# POLAR Investigation of the Sun - POLARIS

T. Appourchaux[1], P. Liewer[2], M. Watt[3], D. Alexander[4], V. Andretta[5], F. Auchère[1], P. D'Arrigo[3], J. Ayon[2], T. Corbard[6], S. Fineschi[5], W. Finsterle[7], L. Floyd[8], G. Garbe[9], L. Gizon[10], D. Hassler[11], L. Harra[12], A. Kosovichev[13], J. Leibacher[1,14], M. Leipold[15], N. Murphy[2], M. Maksimovic[16], V. Martinez-Pillet[17], S.A. Matthews[12], R. Mewaldt[18], D. Moses[19], J. Newmark[19], S. Régnier[20], W. Schmutz[7], D. Socker[19], D. Spadaro[21], M. Stuttard[3], C. Trosseille[1], R. Ulrich[22], M. Velli[2], A. Vourlidas[19], R. Wimmer-Schweingruber[23], T. Zurbuchen[24]

**1)** Institut d'Astrophysique Spatiale, Orsay, France, **2)** Jet Propulsion Laboratory / California Institute of Technology, Pasadena, California, **3)** EADS Astrium, Stevenage, United Kingdom, **4)** Rice University, Houston, Texas, **5)** INAF/Osservatorio Astronomico di Capodimonte, Napoli, Italy, **6)** Observatoire de la Côte d'Azur, Nice, France, **7)** PMOD/WRC, Davos Dorf, Switzerland, **8)** Interferometrics Inc., Herndon, Virginia, **9)** National Space Science and Technology Center, Huntsville, Alabama **10)** Max-Planck-Institut für Sonnensystemforschung, Katlenburg-Lindau, Germany, **11)** Southwest Research Institute, Boulder, Colorado, **12)** Mullard Space Science Laboratory, Holmbury St Mary, United Kingdom, **13)** Stanford University, California, **14)** National Solar Observatory, Tucson, Arizona, **15)** Kayser-Threde GmbH, München, Germany, **16)** Observatoire de Paris-Meudon, France, **17)** Instituto de Astrofísica de Canarias, La Laguna, Spain, **18)** California Institute of Technology, Pasadena, California, **19)** Naval Research Laboratory, Washington DC, **20)** University of St Andrews, United Kingdom, **21)** INAF / Osservatorio Astrofisico di Catania, Italy, **22)** University of California, Los Angeles, California **23)** Christian-Albrechts-Universität, Kiel, Germany, **24)** University of Michigan, Ann Arbor, Michigan

**Abstract**

The POLAR Investigation of the Sun (POLARIS) mission uses a combination of a gravity assist and solar sail propulsion to place a spacecraft in a 0.48 AU circular orbit around the Sun with an inclination of 75° with respect to solar equator. This challenging orbit is made possible by the challenging development of solar sail propulsion. This first extended view of the high-latitude regions of the Sun will enable crucial observations not possible from the ecliptic viewpoint or from *Solar Orbiter*. While *Solar Orbiter* would give the first glimpse of the high latitude magnetic field and flows to probe the solar dynamo, it does not have sufficient viewing of the polar regions to achieve POLARIS's primary objective : *determining the relation between the magnetism and dynamics of the Sun's polar regions and the solar cycle*.

## 1 Introduction

Following the pioneering steps of *Solar Orbiter* (SolO), which will perform the first observations of the Sun out of the ecliptic plane in 2015 – 2023, we propose the POLARIS mission to observe the solar poles for extended periods of time.

The POLARIS mission has been identified in the *Cosmic Vision* document ESA BR-247 under Theme 1: "What are the conditions for planet formation and the emergence of life?" and under Theme 2: "How does the solar system work?" Both themes require a timely implementation of a solar mission observing the poles of the Sun. For Theme 1, under the sub-theme *Life and habitability in the solar system*, POLARIS will map in 3-D the solar magnetic field. For Theme 2, under the sub-theme *From the Sun to the edge of the solar system*, POLARIS will help in understanding the origin of the Sun's magnetic field which requires observations of the field at the visible surface around the poles.

The POLARIS observations will be complementary to those of SolO, which will provide important context for augmenting the POLARIS science. The key feature of the SolO orbit is that at its perihelion of 45 solar radii (0.21 AU), it nearly co-rotates with the Sun allowing longer viewing of the same region of the Sun than from Earth. Like SolO, POLARIS carries both remote sensing and *in-situ* instrumentation, but in addition it provides the long periods of polar viewing time needed to achieve the POLARIS helioseismology objectives. POLARIS will have sufficient viewing of the polar regions to achieve its primary objective: *determining the relation between the magnetism and dynamics of the Sun's polar regions and the solar cycle*.

## 2 Scientific Objectives

Our understanding of the Sun, its corona, and the solar wind has been revolutionized by observations from spacecraft such as SOHO, *Ulysses*, *Yohkoh*, TRACE, RHESSI, *Hinode*, STEREO, and ACE. Yet as we learn more about the Sun from these missions and the complement of ground-based telescopes, the need for information from the polar perspective only increases. The POLARIS mission utilizes a solar sail to place a spacecraft in a 0.48 AU circular orbit around the Sun with an inclination of 75º enabling extended high-latitude studies and direct observation of the solar poles. Observing the polar regions of the Sun with a combination of a Doppler-magnetograph and coronal imagers yields opportunities for major new science in understanding the origin of solar activity. When coupled to total solar irradiance monitoring, UV spectroscopic observations, and *in-situ* particle and field measurements, POLARIS will substantially enhance our understanding of the root causes of solar variability.

The POLARIS mission can address several key science objectives in NASA's strategic plan for Heliophysics. POLARIS, under the name *Solar Polar Imager* (SPI), is included in the intermediate-term mission set for the 2005 Heliophysics strategic roadmap[1].

The mission science objectives of POLARIS are largely derived from those defined in NASA's SPI Vision Mission Study (Liewer *et al.*, 2007 – RD1; Liewer *et al.*, 2008). These objectives which can only be achieved because of the observations enabled by POLARIS short orbital period and highly inclined polar orbit, are as follows:

- *What is the 3D structure of the solar magnetic field, and how does it vary over a solar cycle?*
- *What is the 3D structure of convection and circulation flows below the surface, and how does it affect solar activity?*
- *How are variations in the solar wind linked to the Sun at all latitudes?*
- *How are solar energetic particles accelerated and transported in radius and latitude?*
- *How does the total solar irradiance vary with latitude?*
- *What advantages does the polar perspective provide for space-weather prediction?*

Each of above mission objectives is discussed in detail below. A summary of POLARIS science objectives and the contributions of the different instruments in the payload are given in Table 1. The POLARIS mission will also contribute to important "multi-viewpoint" science objectives (*e.g.* many of NASA's STEREO and *Sentinel* science objectives) as well as complement near-Earth remote sensing and *in-situ* measurements (*e.g.* SDO, *Hinode*, and successors). However, we have limited the primary objectives to those that require the viewing geometry provided by the POLARIS orbit.

## *2.1 What is the 3D Structure of the Solar Magnetic Field and How does it vary over a Solar Cycle?*

Our understanding of the structure and dynamics of the solar magnetic field has grown significantly with the advances provided by SOHO, *Ulysses*, *Yohkoh*, and TRACE. STEREO, *Hinode*, SolO, and SDO will further advance our knowledge in several key and complementary areas relating to the physics governing solar variability. However, basic questions about the strength of the polar magnetic field, the distribution and evolution of magnetic fields in the polar coronal holes and the evolutions and diffusion of the high latitude magnetic fields remains unanswered. Such questions can only be fully explored from extended observations from out of the ecliptic plane. POLARIS carries a synergistic complement of remote-sensing instruments enabling us to determine the 3D structure of the solar magnetic field and how it varies over a solar cycle. To get the full benefit of the polar viewpoint in understanding the structure and dynamics of the solar magnetic field, the POLARIS payload includes a magnetograph and Doppler imager, a white-light coronagraph, an EUV imager, sensitive to the hot, million-degree lower corona and the cooler upper chromosphere, and a UV spectrograph, to measure outflows in the chromosphere, transition region, and corona.

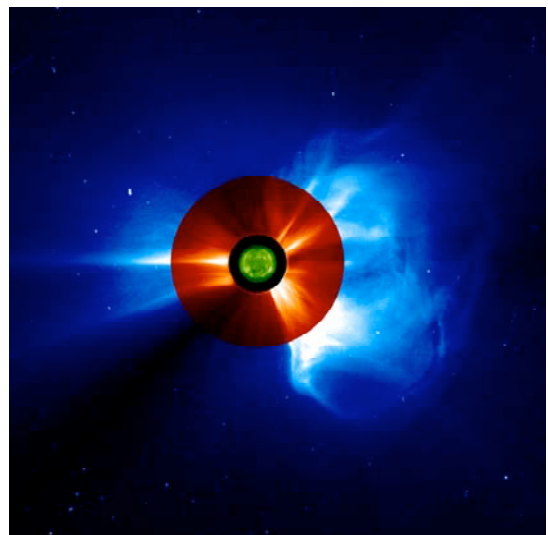

Figure 1: Composite view of the corona and the Sun taken by EIT and LASCO aboard SOHO, obtained with the FESTIVAL tool.

---

[1] see web page: sec.gsfc.nasa.gov/sec_roadmap.htm

Magnetograms, taken every 5 minutes, will be used to study the evolution of active regions, flux transport, and the solar cycle field reversal, with emphasis on the polar regions. The POLARIS coronagraph will provide high-latitude views of the extended corona from $R = 1.5 - 15\ R_{Sun}$. These observations will show the impact of a CME on the global coronal magnetic field, how the corona recovers from these disruptions, and how the streamer belt ultimately reforms (Figure 1). We will be able to measure the longitudinal extent of CMEs and, when nearly simultaneous CMEs appear on opposite sides of the Sun as viewed from Earth, the polar view point will make it possible to determine whether one event triggered the other or whether they are, in fact, part of the same nearly global CME. At solar minimum, the POLARIS polar view of the streamer belt will determine whether the streamer belt is uniform in longitude (see, for example Wang *et al.*, 2007), or whether it is filamentary (Liewer *et al.*, 2001).

Using the coronal imagers on POLARIS, we can observe the initiation and evolution of Earth-directed CMEs and determine the spatial relation between CMEs and their coronal sources. The POLARIS coronagraph and EUV imager will observe polar plumes and small-scale structures in the polar coronal-hole region. We will also better determine coronal-hole boundary locations for refining global field topology using the EUV imager's 304-Å line.

The UV Spectrograph will provide detailed plasma diagnostics of the coronal structures seen by the EUV imager and will allow comparison of the magnetic-field geometry observed by the magnetograph to the detailed chromospheric and transition-region velocity structures. In particular, the UV Spectrograph will measure directly the outflows of CMEs and polar plumes as well as the fast solar-wind outflows coming from polar and equatorial coronal hole regions.

The "accurate" heliographic angles (from line of sight) of the magnetograph is about 0°-60°; beyond this, foreshortening causes the signal-to-noise ratio to drop, leading to larger uncertainties in the magnetic field determinations. With observations from both POLARIS and Earth, each point on the surface in the active region belt is visible for about 50% of the time in the 60° "accurate" heliographic angles, thus allowing improved studies of the evolution of active regions in response to the flows determined by helioseismology.

POLARIS will compare the very best 3D MHD models of the corona with the very best global observations. The magnetic-field boundary conditions for the models depend on the surface magnetic field, which is presently not measured at high latitude and which is constructed for a full Carrington rotation with data that is up to 26 days old. POLARIS will provide these detailed high-latitude and polar magnetic-field boundary conditions which, when combined with other magnetograph observations from the Earth, will yield almost complete latitudinal coverage of the surface magnetic field. Moreover, POLARIS will yield a full 360º of data for one hemisphere down to ≈30º latitude to fill in areas where the field is rapidly changing and reduce the length of time over which the data needs to be accumulated or assumed (*e.g.* Schrijver and DeRosa, 2003).

Combined observations from the POLARIS four remote sensing instruments and by near-Earth observations, together with coronal models using more accurate magnetic boundary conditions, will improve studies of how the coronal magnetic field responds to changes in the convective flows and photospheric fields on both active region and global spatial scales. POLARIS will enable the exploration of the formation, evolution, and demise of solar structures and will provide key information on active region heating and global field connectivity. POLARIS will enable us to follow the long-term evolution of polar coronal holes, which contribute to our understanding of the cyclical changes of the solar magnetic field and how the poleward migration of field influences the fast solar wind as well as the global corona.

### *2.2 What is the 3D Structure of Convection and Circulation Flows Below the Surface, and How does it affect Solar Activity?*

This objective addresses directly one of the fundamental questions in solar physics: *How and where does the solar dynamo operate, and in what way do the fields created by the dynamo move up through the visible surface?"* This requires knowledge of the 3D structure of the flows in the solar interior, which in turn requires observations of the polar regions of the Sun.

Doppler data from the *Michelson-Doppler-Imager* (MDI) on SOHO and the ground-based *Global Oscillation Network Group* (GONG) have revolutionized our views of the structure and dynamics of the convective region and the solar dynamo (Schou et al., 1998; Kosovichev et al., 2000). Figure 2 shows measurements of the differential rotation as a function of depth from SOHO/MDI data. A region of large velocity shear can be seen at about 0.7 $R_{Sun}$; it is this shear, at the base of the convection zone (tachocline) that is now thought to drive the large-scale solar dynamo. The current ecliptic-viewpoint observations do not provide measurements of sufficient accuracy for helioseismic inversion of the solar structure and rotation in the polar regions. At present, the internal structure and the differential rotation have been measured with great precision using global helioseismology techniques, except in the polar regions (75° – 90°) and in the energy-generating core. POLARIS will enable measurements in both regions by using a new technique called stereoscopic helioseismology based on time-distance helioseismology. This technique is to use observations from helioseismic instruments placed at least 120° from each other: one aboard POLARIS and one ground-based instrument such as GONG, or space-based instrument such as HMI aboard Solar Dynamic Observatory. The observation of the propagation of waves in the core of the Sun from both sides will enable to recover the structure of the core.

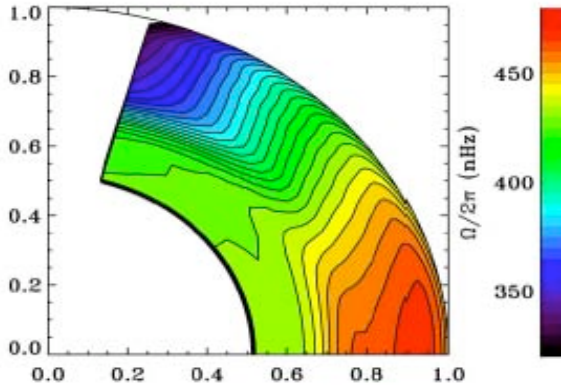

Figure 2: Solar rotation rate ($\Omega/2\pi$ nHz) versus depth (solar surface is at 1.0). The white polar and deep regions where the rotation rate is currently unknown will be filled in using POLARIS (Based on Schou et al., 1998).

In spite of the enormous progress that has been made in helioseismology, we still have only a limited understanding of the relationship between the convective flows, the solar cycle, and the solar dynamo, in part because of the limitations of ecliptic-viewpoint solar observations. Critical processes in the polar regions, such as magnetic flux transport and magnetic-field polarity reversals, which define the strength and duration of the solar cycle, are consequently poorly understood. So far, the meridional flows have been measured only for relatively low latitudes, up to about 60° (Haber et al., 2002; Zhao and Kosovichev, 2004). Additionally, the torsional oscillations, which are extremely important because active regions tend to emerge in the shear layers between faster and slower streams, have not been measured reliably at high latitudes due to foreshortening and low signal-to-noise ratio of the solar oscillation signal near the limb.

Helioseismology requires long, nearly-continuous observing periods at a high temporal cadence in order to resolve the various spatial and temporal scales associated with processes of generation and dissipation of solar magnetic fields. These scales range from supergranulation cells to the global meridional and zonal flows. Supergranulation defines the magnetic network with a typical spatial scale of 30 Mm and lifetime of about several days to a few weeks, as well as the larger-scale dynamics of active regions occupying 50 – 100 Mm and evolving on the scale of weeks to a couple of months. The meridional and zonal flows occupy the whole convection zone 200 Mm deep and vary with the solar cycle on the time scale of a year.

The high-latitude viewing will provide a unique opportunity to study the dynamics of meridional flows and rotation in the polar regions and search for deep longitudinal structures in the tachocline via local helioseismology. Figure 3 shows the sound wave ray paths (red lines) that will enable us to measure the subsurface meridional flows (solid lines) using the POLARIS high-latitude viewpoints.

The high-latitude orbit will allow us to obtain better coverage of the deep polar regions using the observing scheme with two vantage points, POLARIS and an ecliptic-based helioseismology instrument. Two vantage points widely separated in longitude and latitude allow helioseismology to probe ray path penetrating deep into the Sun, down to the tachocline, by correlating the signals from the two view points.

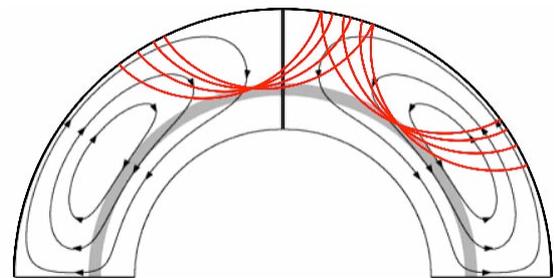

Figure 3: Closed circulation lines: Possible streamlines of meridional flow. Gray and red lines: Sound wave (p mode) ray paths that can be used to measure the subsurface meridional flows using POLARIS high latitude.

## *2.3 How are Variations in the Solar Wind Linked to the Sun at all Latitudes?*

*In-situ* measurements of the solar-wind plasma, the heliospheric magnetic field, energetic particles, and isotopic and elemental composition in the polar regions will link variations in the high-latitude heliosphere to solar-surface conditions. The *Ulysses* mission has so far provided only two 360º latitude scans at varying radius (the *Ulysses* 6-year orbit extends to Jupiter). POLARIS, with its 4-month orbital period, will produce three 360º rapid latitude scans *per year* at a constant radius. The rapid latitude scan and closeness to the Sun will enable unprecedented observations of the evolution of the solar wind from its source to the spacecraft with minimal effects from stream-stream interactions. Previous work (e.g. (Neugebauer *et al.*, 2002; Liewer *et al.,* 2004) has shown that *in-situ* measurements of the solar wind magnetic field, plasma velocity and ion composition are needed to trace the solar wind to its solar source. Coupled with the more accurate magnetic-field extrapolations expected as a result of POLARIS, the rapid latitude scans will more accurately determine the source regions of the solar wind and determine how variations in the solar wind are linked to the Sun at all latitudes.

The connection between *in-situ* solar wind and its solar sources can be analyzed with measurements of solar-wind speed, energy flux, magnetic field, chemical composition, and ionization state. Compared to other missions, POLARIS will have significantly improved temporal resolution, a coronagraph, an EUV Imager, and the UV Spectrograph which will provide complementary observations to those from the magnetometer and solar wind analysers. Observations of the solar wind at high latitude near 0.48 AU may allow the connection between the plumes and solar wind features to be established (De Pontieu, 2007).

Determining the physics and structure of the wind's acceleration requires tracing the outflowing plasma from the solar surface, through the chromosphere and transition region, to the corona. Imaging spectroscopic studies using SOHO/SUMER have demonstrated that Dopplergrams of chromospheric and coronal spectral lines can trace the origins of the solar wind (Hassler *et al.*, 1999). Full-disk synoptic Dopplergrams from the POLARIS UV spectrograph will image the size, shape, and velocity structure of the solar wind coming from coronal holes as a function of time, providing critical boundary conditions for *coronal and solar wind models*. Radial velocities in polar coronal holes range from 5 – 12 km/s, whereas radial velocities in the equatorial coronal holes are significantly lower with values on the order of 3 – 8 km/s (Buchlin and Hassler, 2000). POLARIS will explore this latitudinal variation on a rotational and solar-cycle time scale for the first time.

It has long been established that most abundance fractionation processes have their roots in the chromosphere (*e.g.* Geiss, 1982). Therefore, the patterns of abundance variations, found both in the solar wind and in the corona (measured from *in-situ* and EUV spectroscopic observations), have already taken place and remain roughly constant in the observed structures, with the exception perhaps of those structures that are magnetically-confined and long-lived. Such an advantage of elemental composition analysis is further reinforced by out-of-the-ecliptic, relatively rapid latitudinal scans. In particular, the severe problems affecting near-ecliptic measurements, induced by the overlap of numerous structures along the line of sight (*e.g.* Ko *et al.*, 2007), would be significantly reduced when a polar view becomes available.

## *2.4 How are Solar Energetic Particles Accelerated and Transported in Radius and Latitude?*

The POLARIS orbit greatly improves our ability to explore energetic particles in the inner heliosphere and to determine how particles are accelerated and transported in radius and latitude. How are particles accelerated in the largest solar particle events? For large events, CME shock acceleration is occasionally seen to continue to energies >10 MeV/nucleon at 1 AU. At 0.48 AU, POLARIS will be within the prime shock acceleration region more often and over a much broader energy range, permitting detailed tests of theory with observations of particle spectra, shock structure, and magnetic fluctuations and solar radio emissions.

Understanding the relative roles of CME-driven shocks versus flare-associated processes in solar energetic particle (SEP) acceleration is an important issue. POLARIS will address the important question of where solar particle events originate. Figure 4 shows how the velocity dispersion of the energetic particles can be used to determine the SEP injection time, which can then be compared to the flare and CME onset times to relate the SEP events to their source. It will be possible to relate flare-accelerated particles to their sources more accurately and to identify the time and altitude at which CME-driven shock acceleration begins with much greater precision. POLARIS will also carry the instruments necessary to determine the flare (EUV telescope) and CME (coronagraph) onset times as well as the CME height as a function of time. The radio instrument and energetic particle packages on POLARIS will provide details about the CME shock and the associated electron acceleration that can provide important information about the subsequent evolution of the SEP event (Posner, 2006).

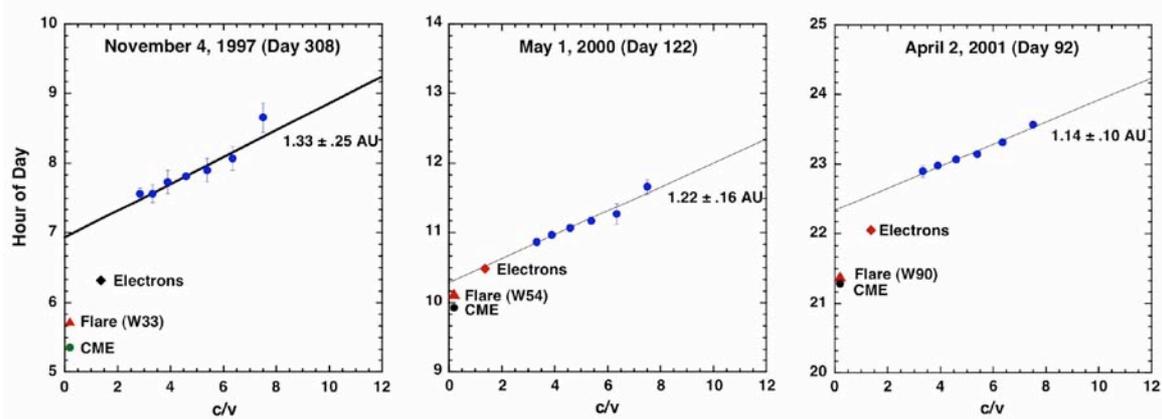

Figure 4: Time of arrival of energetic particles versus inverse velocity for 3 SEP events. Extrapolation to intersection shows particle injection time for comparison with flare and CME onset times (Mewaldt *et al*, 2003).

POLARIS is an ideal platform to more closely examine the rapid transport of energetic particles in latitude during CMEs with its magnetometer and particle instruments. *Ulysses* observations have shown that the same solar particle events observed at Earth are also observed at high latitudes (Dalla *et al*., 2003). POLARIS will make 6 rapid latitude scans a year, permitting frequent snapshots of the latitudinal gradients in the intensity of anomalous and galactic cosmic rays which can be used to determine the diffusion coefficients for particle transport both parallel and perpendicular to the magnetic field. These measurements, complemented by better magnetic field models (using the extended coverage magnetograms made possible by POLARIS), will dramatically increase our understanding of energetic particle transport in latitude.

POLARIS will allow extensive measurement of solar type II and type III radio emissions from various radial and latitudinal locations. These radio emissions are produced by subrelativistic electrons, that are either ejected from active regions and travelling outward along open magnetic field lines (type III) or accelerated by CME associated shocks (type II). POLARIS will provide decametric-to-hectometric radio spectroscopy to trace electron propagation through space, identify the magnetic connectivity of the spacecraft to solar sources, and determine timing of electron release in the corona.

POLARIS will also allow the study of the intense *in-situ* Langmuir-like waves that are frequently observed in the solar wind with the above mentioned type II and type III radio emissions. The physics of the plasma/beam interaction producing these Langmuir waves is still in debate as are the physics of the conversion of these waves into electromagnetic radio emissions. The radio waves can be observed remotely, indicating Langmuir wave occurrence nearer to the sun (at higher frequencies corresponding to higher solar wind densities). Because of the specific POLARIS trajectory, it will make important progress in this field by observing these particle/waves interactions at various radial and latitudinal locations in the inner heliosphere.

## 2.5 How does Total Solar Irradiance Vary with Latitude?

The primary motivation for monitoring the total solar irradiance variability is to determine solar influence on global change. However, the total solar irradiance (TSI) and its temporal variation are crucial to our understanding not only of the Sun-Earth Connection but also of the Sun as a star. TSI is the solar radiation in a given direction, integrated over all wavelengths and

- 6 -

over the whole solar disk, in units of power per unit area (*e.g.* W/m$^2$). The Earth-science community needs these observations to provide the solar energy input to the Earth's climate system. TSI has been measured continuously from space since 1978 by a number of experiments, but always from the ecliptic, near-Earth viewpoint (Fröhlich and Lean, 2004).

The POLARIS orbit will enable us to determine how the total solar irradiance varies with latitude. POLARIS provides the necessary viewing geometry from which to characterize the various contributions to the Sun's radiative output at all latitudes. The TSI varies with solar activity on a range of time scales, including periodic variations associated with solar rotation ($\approx$27 days) and the solar cycle ($\approx$11 years). The solar cycle TSI modulation is known to be about 0.1%, with larger values occurring at solar maximum, largely because the contributions of the bright faculae and network more than compensate for the reduction due to the dark sunspots. The TSI also varies as a result of the emergence and evolution of active regions containing spots and faculae on intermediate time scales (*i.e.* months). During the passage of a large spot group across the centre of the solar disk, the TSI can vary by as much as 0.3% over these few days.

With more complete coverage of the TSI in latitude, POLARIS can address questions relating to the mechanism of the variation in solar luminosity. Moreover, measurements of the TSI from the different perspectives provided by POLARIS will aid in our interpretation of the irradiance measurements of Sun-like stars whose polar orientations are unknown and whose activity cycles are apparently quite different. Long-term observations of these stars are important for establishing possible future behaviours of the Sun. POLARIS may help answer why solar irradiance variability is $\approx$1/3 of the variation observed in other Sun-like stars.

The disk-integrated TSI measurements will be complemented by spatially resolved intensitygrams from the Doppler-Magnetograph. POLARIS will also explore the variations in the Sun's UV and EUV spectral irradiance with the UV/EUV spectrograph. These observations will constrain models of the TSI based on radiative contributions from various solar surface features (sunspots, faculae, magnetic network and quiet Sun) and for the first time TSI models will be constrained by polar observations.

## *2.6 What Advantages does the Polar Perspective Provide for Space Weather Prediction?*

Developing a predictive capability for the dynamic space environment will be important for both human and robotic explorers. The polar viewpoint will provide unique space weather observations that will augment data from other spacecraft deployed in the inner heliosphere as well as solar observatories on Earth. POLARIS can determine the advantages of the polar perspective for space weather prediction. The POLARIS coronagraph will monitor Earth- (and Mars-) directed CMEs from their point of origin to 15 R$_{Sun}$ from the high-latitude perspective. For events which are "halos" as viewed from Earth, this will give far better speed estimates so that the very fast CMEs those associated with the largest and most hazardous solar energetic particle events can be identified. The *in-situ* instruments on POLARIS will provide crucial additional information about the speed of the interplanetary CME and its associated shock through measurements made with the radio instrument, while the magnetometer can give the orientation of the magnetic field within the ejecta. With the limitations of SOHO's longitudinal coverage, limb CMEs and associated SEPS are frequently observed when their source regions are still behind the limb. The increased solar coverage in longitude and latitude will yield more frequent observations of the active region sources of CMEs and SEPs, further enabling more reliable space weather forecasts.

Helioseismic observations of subsurface flows will give advanced warning of the emergence of large amounts of flux which may eventually help us better predict whether potentially dangerous active regions come around the limb. An important result of this will be the ability to increase the safety and productivity of astronauts.

As with Earth weather prediction, space weather prediction will rely heavily on numerical models. The POLARIS observations, complemented by near-Earth observations, will provide more accurate information on the global magnetic field (given the greatly improved solar coverage in both longitude and latitude) resulting in greatly improved predictive models. Especially valuable will be the POLARIS unique measurements of the polar magnetic fields. The improved knowledge of the photospheric magnetic boundary condition will lead to more accurate potential and MHD models of the coronal and heliospheric magnetic fields as discussed in Section 2.1. Given these advantages, a polar mission may well be a crucial component of a 3D array of heliospheric monitoring spacecraft, providing operational support for Space Weather forecasting, and a more complete coverage of the solar surface.

Given that POLARIS will often be in a position to provide early warning of space weather events affecting Earth, we plan to study the possible implementation of a Beacon Mode on POLARIS along the lines of that currently in operation on STEREO.

## 3 Payload

### *3.1 Overview of Proposed Payload*

The payload will consist of remote sensing and *in-situ* instruments. The remote-sensing instruments are:

- Dopplergraph and magnetograph imager
- White-light coronagraph
- EUV imager
- UV spectrograph
- Total Solar Irradiance monitor

All imaging instruments will have a typical pixel resolution of 2 to 4 arcseconds. The *in-situ* instruments are:

- Magnetometer
- Solar-wind ion composition and electron spectrometer
- Energetic particle package
- Radio and plasma wave package

Table 1 summarizes the scientific objectives that are addressed by each instrument, while Table 2 sums up the required resources. All remote instruments shall be no longer than one meter.

**Table 1: Science Objectives and Contributing Instruments**

| Scientific Objectives | Remote sensing | | | | | *In situ* | | | |
|---|---|---|---|---|---|---|---|---|---|
| | DSI | COR | EUVI | TSI | UVS | MAG | SW | EPP | RPW |
| What is the 3D structure of the solar magnetic field, and how does it vary over a solar cycle? | × | × | × | × | | × | | | |
| What is the 3D Structure of Convection and Circulation Flows Below the Surface, and How does it affect Solar Activity? | × | × | × | × | × | | | | |
| How are variations in the solar wind linked to the Sun at all latitudes? | | × | × | | × | × | × | × | |
| How are solar energetic particles accelerated and transported in radius and latitude? | × | × | × | | | × | × | × | × |
| How does the spectral and total solar irradiance vary with latitude? | × | | × | × | × | | | | |
| What advantages does the polar perspective provide for space weather prediction? | × | × | × | | × | × | × | × | × |

COR: Coronagraph  
TSI: Total Solar Irradiance Monitor  
EUVI: Extreme Ultra-Violet Imager  
DSI: Doppler-Stokes Imager  
UVS: Ultra-Violet Spectrograph  

MAG: Magnetometer  
SW: Solar Wind Ion Comp. and Electron Spec.  
EPP: Energetic Particles Package (20 keV – 10 MeV).  
RPW: Radio and Plasma Waves  

Table 2: Payload Resource Summary

| | Remote Sensing Instruments | | | | | In-situ Instruments | | | |
|---|---|---|---|---|---|---|---|---|---|
| | DSI | COR | EUVI | TSI | UVS | MAG | SW | EPP | RPW |
| **Mass (kg)** | 25 | 10 | 10 | 7 | 15 | 1.5 | 10 | 9 | 10 |
| **Power (W)** | 37 | 15 | 12 | 14 | 22 | 2.5 | 15 | 9 | 15 |
| **Absolute pointing (3 σ)** | 30′ | 10″ | 30′ | 30′ | 0.5′ | N/A | 120′ | N/A | N/A |
| **Pointing stability (3 σ)** | 0.2″ in 10 s | 7″ in 1s | 1″ in 10 s | 6′ in 1 s | N/A | N/A | 6′ in 1 s | N/A | N/A |
| **Data rate (kbps)** | 75 | 40 | 40 | 0.4 | 10 | 0.6 | 0.2 | 1.0 | 5 |
| **FOV** | 1.5°×1.5° | 16° | 1.5° | 2.5° | 10″ × 1.4° | N/A | 20° × 160° | See text | 4 π |

Outlined in grey: the instruments requiring internal pointing

## 3.2 Payload description

### 3.2.1 Doppler and Stokes Imager

The DSI Doppler images will be used to perform local-helioseismic imaging of sub-surface structures and flows in the polar regions, and will perform stereoscopic helioseismic imaging of the polar regions and the deep interior in conjunction with ground-based instruments such as the Global Oscillation Network Group (GONG) or near-Earth spacecraft.

The Doppler and Stokes imager will provide solar radial velocity every 45 seconds (in order to resolve the helioseismology oscillations), with a pixel resolution of 4 arcseconds, and vector magnetic-field maps of the whole Sun every five minutes with a pixel resolution of 2 arcseconds. The entrance aperture will be about a 6-cm diameter, sufficiently large to provide the required spatial resolution and sensitivity (15 m/s per pixel). The instrument will have a full field of view of 1.5° × 1.5°. The focal plane detector will be a 2K × 2K active-pixel sensor (APS). Radial velocity images will be binned to 1K × 1K or 512 × 512, depending on the type of helioseismic objectives. The Stokes images will be at full resolution. Radial velocity images require an image stability of better than 1/20 of a binned pixel (0.2 arcseconds at 3 s). A tip-tilt mirror will provide this pointing stability during the exposure.

The data rate will be 75 kbps allowing the downlink of velocity images every 45 seconds and three Stokes-parameter images every five minutes; using a lossless compression factor of two, each parameter will encoded at 4 bits/pixel. The instrument will have two main operational modes: calibration and observation. The calibration mode will be used to calibrate on-board the measured solar radial velocities.

A solar line sensitive to magnetic field, such as Fe I 6173 Å, will be chosen for measuring solar radial velocities and magnetic field. The access to a large variety of wavelengths is made possible by the availability of the various concepts of spectral analyzers such as Michelson and Lyot filters, Fabry-Perot etalons or Magneto-optical filters. Michelson and Lyot filters have been used in several space instruments such as the Michelson and Doppler Imager (MDI) aboard SOHO (Scherrer *et al.*, 1995), and will be used by the Helioseismology and Magnetic Imager (HMI) aboard SDO (Scherrer *et al.*, 2006). Fabry-Perot etalons with piezoelectric spacers have been flown aboard the UARS mission such as the HRDI instrument (Skinner *et al.*, 2003); they are also developed as back-up solutions for SolO (Figure 5). Solid etalons based on lithium niobate have been also developed for ground-based applications, and are being developed for balloon-borne and space applications such as for ESA's SolO (Alvarez-Herrero *et al.*, 2006; Trosseille *et al.*, 2007). Magneto-optical filters based on sodium and potassium cells have been used for ground-based applications for the past twenty years (Tomczyk *et al.*, 1995) and instruments using other spectral lines have also been developed (Murphy *et al*, 2005). Resonance cells of a similar concept have been operating in space aboard SOHO on the GOLF instrument (Gabriel *et al.*, 1995) for more than 10 years.

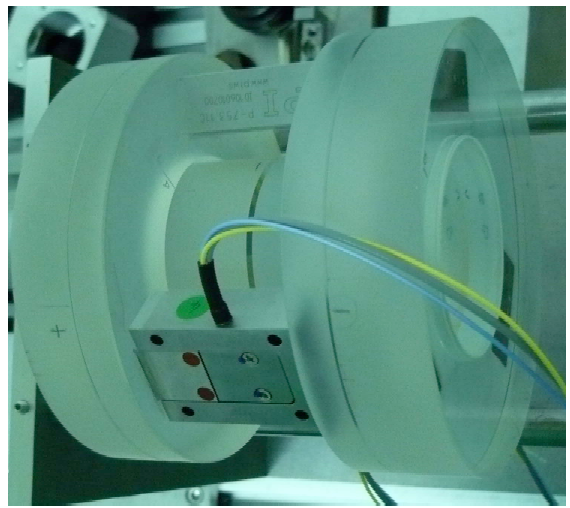

Figure 5: Fabry-Perot etalon developed by Winlight, France as a prototype for SolO (Trosseille, 2007).

### 3.2.2 Coronagraph

Coronagraphs occult the bright solar disk to observe the extended faint corona through the visible light emitted by the Thomson scattering of the photospheric radiation by the coronal electrons. The design is actually a shortened version of the COR2 coronagraphs constructed for the STEREO mission (Figure 6). The optical design has been modified to incorporate a 5-element external occulter and to achieve good stray light performance over a 16º field of view with a 60-cm long instrument. The focal plane array incorporates the same 2K × 2K CCD camera and electronics used in the COR2 instruments. The coronagraph will observe the corona from 1.5 to 15 solar radii with 28-arcsec pixels. The coronagraph has polarimetric capability with the inclusion of a rotating polarizer in the beam. The total mass of the instrument is 10 kg, and its power consumption is 15 W. To achieve the required stray-light rejection levels, the coronagraph requires pointing stability of 10 arcseconds (3-σ) during the exposure.

To satisfy the mission objectives, the coronagraph must provide total and polarized brightness (*B* and *pB*, respectively) images of the corona up to $10 - 15$ $R_{Sun}$. Both total *B* and *pB* images can be derived from a set of 3 images each taken at polarization angles 60º apart (Billings, 1966). A cadence of 12 minutes for each 3-image set is sufficient to follow the evolution of CMEs and examine their azimuthal structure and that of the corona. It is also sufficient to provide space-weather information for Earth- and Mars-directed ejections. The resulting average data rate is 40 kbps, assuming 2:1 data compression. For investigations of SEP production at CME shocks, a faster observing sequence could be adopted within the same telemetry allocations. It is achieved by onboard summing of 2 images (taken at polarization angles 90º apart), thus producing a single total *B* image every 4 minutes.

### 3.2.3 EUV Imager

The EUV imager proposed for the POLARIS mission derives its heritage from the SolO EUI/FSI instruments (Figure 7). It is based upon a simple Herschelian telescope with a single off-axis mirror and a metallic filter (aluminium or aluminium/zirconium). The entrance aperture is about 15-mm diameter. The detector is a 2K × 2K Active Pixel Sensor (APS). There are 2 such telescopes, each mirror being coated with different multilayer dielectrics. This allows simultaneous observations in 2 relatively narrow passbands. We have baselined the 2 lines centred at 195 and 304 Å because they capture the majority of events that occur in the chromosphere, in the transition region and in the corona. Furthermore, the multilayer coatings for both passbands are well understood, are easy to produce and carry a long flight heritage on SOHO/EIT and STEREO/SECCHI. A single-shot door is considered along with a shutter and a filter wheel. This design allows full-disk observations of the corona out to 1.4 $R_{Sun}$ with 2.6 arcsecond pixels, similar to EIT images. This corresponds to a resolution on the Sun twice that of SOHO/EIT and slightly better than that of STEREO/EUVI. The imager has very modest pointing requirements but requires one arcsecond (3 σ) pointing stability over a typical 10-second exposure (195Å). A Guide Telescope (GT) such as used by TRACE and STEREO, will be used for providing an active pointing to the instrument.

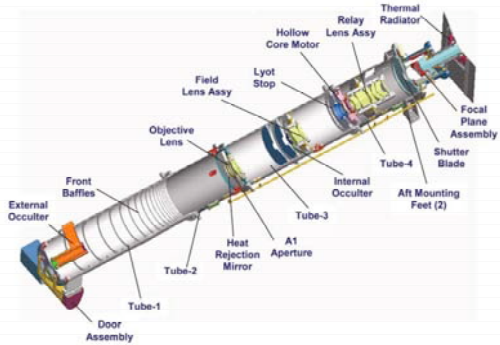

Figure 6: Exploded view of the COR2 of STEREO / SECCHI

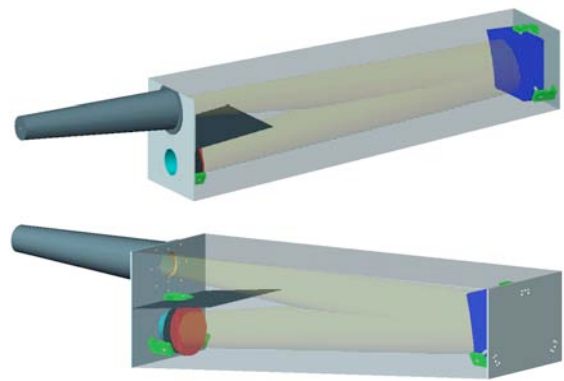

Figure 7: Front and back view of one of the EUVI telescope. The dimensions of the instrument fit into $80 \times 17 \times 17$ cm$^3$. The GT (not shown here) moves the mirror (in blue)

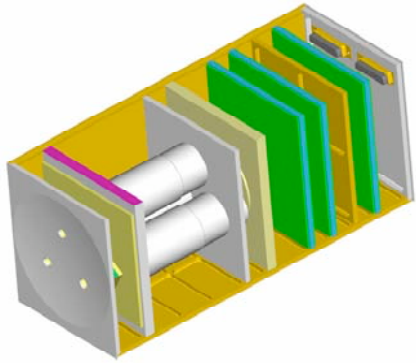

Figure 8: View of the POLARIS TSI Monitor with the top and side covers removed. Sunlight enters the instrument through the 3 apertures in the front plate and will be absorbed in the receiver cavities at the back of the cylindrical housings. The electronics are located in the back of the package (green).

The EUV imager data products are full-disk images of the corona in the 2 wavelengths discussed. The default observing sequence will be an image in each wavelength taken simultaneously, every 8 to 10 minutes. This observing program fits well in the telemetry allocation of 40 kbps (10-minute cadence, 2:1 compression) with sufficient cadence to support most of the mission objectives. The change to the rapid cadence program could easily be enabled by an onboard flare flag.

**3.2.4 Total Solar Irradiance Monitor**

The POLARIS TSI Monitor is a new-generation, room-temperature absolute radiometer based on the electrical substitution principle (ESR). It employs 3 cavities that are all pointed towards the Sun but can be shaded individually. The cavities are operated differentially: one cavity (the reference cavity) remains shaded and heated with a constant electrical power while the electrical heating power in the remaining two cavities (active cavities) is actively controlled to maintain equal heat fluxes from all 3 cavities to the heat sink. While one of the active cavities will be alternately exposed and shaded every 10 seconds to provide measurements of the TSI, the backup cavity will be exposed only once per month to monitor the degradation of the blackening of the measuring cavity. All 3 cavities are capable of measuring solar irradiance and can be assigned for reference, measuring, or backup. The optical design of the POLARIS/TSI Monitor will have 3 precision entrance apertures of 4 mm nominal diameter in the front plate of the instrument. The optical design is such that all light entering the front apertures will be absorbed in the cavities, thus reducing problems with stray-light and thermal oscillations in the precision aperture and light baffle. For the proposed 0.48 AU orbit the dissipated heat per cavity will be held constant at 100 mW throughout the mission.

The triple-cavity design offers redundancy as well as the possibility for in-flight monitoring of the instrumental performance and degradation (Figure 8).

The data will be evaluated by a frequency analysis of the instrument's response by extracting the signal at the fundamental of the shutter frequency. This so-called phase-sensitive mode of operation offers many advantages over the normal active-cavity operation in *e.g.* SOHO/VIRGO, which relies on reaching an equilibrium state before measurements can be taken: for example it reduces the effect of the non-equivalence and is insensitive to any out-of-frequency or out-of-phase noise.

**3.2.5 UV Spectrograph**

A UV spectrograph measures line-of-sight velocities from the observed Doppler shift in the wavelength of a given spectral emission line formed at a particular height and temperature in the atmosphere. The POLARIS/UV Spectrograph is based on a miniaturized version of the *Rapid Acquisition Imaging Spectrograph* (RAISE) spectrograph (scheduled to be tested on a sounding rocket), but scaled down to the size of the *Rosetta*/ALICE UV spectrograph, with a total length of 600 mm and mass of 15 kg. RAISE is serving as a prototype for a next-generation imaging spectrograph, which provides high-cadence stigmatic imaging at 2 spectral passbands simultaneously. The spectrograph includes a single-element off-axis parabola mirror and a toroidal variable line space (TVLS) grating with an instantaneous field of view of the entrance slit of 10 arcseconds × 85 arcminutes to image an entire chord across the full solar disk at 0.48 AU, with a spatial resolution of 10 arcseconds, corresponding to 3700 km on the Sun (Figure 9). With this design, a single raster scan can produce a full-Sun spectroheliogram in 17 minutes. The baseline design we have chosen covers the 1166Å – 1266Å wavelength range, which is sufficient to enable sampling of the full range of heights and temperatures in the solar atmosphere, from the chromosphere, through the transition region and into the million-degree corona.

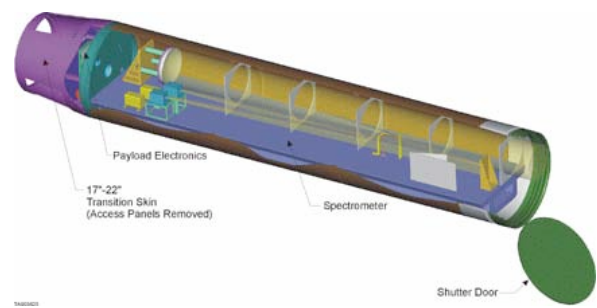

Figure 9: View of the RAISE spectrograph as designed for a sounding rocket.

The data products include 3-D spectral images (spectroheliograms) of the Sun (two spatial dimensions, and one spectral dimension) with intensity, line-of-sight velocity, and non-thermal temperature as the parameters in the spectral dimension. The data products would be similar to that of a magnetogram, with Doppler velocity replacing that of magnetic field strength.

### 3.2.6 Magnetometer

Low-frequency magnetic-field measurements on POLARIS will be made using a laser-pumped vector helium magnetometer (LPVHM). The LPVHM combines the heritage and stability of the vector helium magnetometer, flown on *Ulysses* and *Cassini* for example, but has a significantly lower mass. Much of the mass reduction in the LPVHM is achieved by eliminating the pumping lamp from the sensor and replacing it with a diode laser that is coupled from the magnetometer electronics package within the spacecraft via an optical fibre. The return signal from the sensor is also fibre coupled to a detector located within the spacecraft electronics. The use of a laser to replace the lamp also allows a significantly higher instrument sensitivity, which in this case is traded for a reduction in sensor volume. This results in a simplified sensor, without a pumping lamp or detector, which weighs 150 g. Accommodation of the instrument is further simplified by replacing the detector signal and lamp power cables with optical fibres. Instrument electronics are based on heritage designs and will interface with the spacecraft DPU, which will provide commanding and data handling for the instrument. The instrument electronics weigh 1 kg, giving a total instrument mass of 1.3 kg.

Two LPVHM sensors are mounted on a 5 m-magnetometer boom to ensure that the residual spacecraft field at the outboard sensor is < 0.2nT. Two sensors are used, separated by 1.5 m, both for redundancy and to provide an independent estimate of spacecraft field. The instrument will collect data at a rate of 10 vectors/second, which leads to a peak data rate of 576 bps, including instrument housekeeping data.

### 3.2.7 Solar Wind Ion Composition and Electron Spectrometer

This instrument consists of 3 sensor heads, the solar wind ion sensor (SWI), and two identical solar wind electron sensors (SWE). Both sensors have ample heritage, but the POLARIS design takes advantage of novel electronics advances to reduce mass for accommodation into the mission. The electron sensors are qualitatively similar to the sensors currently in flight on FAST and THEMIS.

The two SWE sensors each have a look direction of $180^{o}$ in azimuth, split in 8, $22.5^{o}$ bins (one per anode) with a sweeping voltage bending the field-of-view for $90^{o}$ elevation in $22.5^{o}$ bins. Measurements over $4\pi$ sr, are made once every 5 seconds, but lower data rates are typically transmitted due to telemetry constraints. Each analyzer consists of a top-hat (~$90^{o}$ symmetric quadraspherical) ESA that provides a ≈180° × 10° fan-shaped field of view. The combined mass and power of the two SWE sensors are 2 kg and 3 W. The energy range is 0.005 keV with 15% resolution.

The SWI sensor consists of two subsystems, the electrostatic analyzer and the Time-of-flight/energy (TOF/E) assembly, to enable rudimentary ion composition relevant for the analysis of the solar wind origin in the corona (Neugebauer *et al.*, 2002). The TOF/E system is at a voltage of 10 kV as compared to 15 kV in FIPS / MESSENGER. The principle of operation is as follows: ions are separated according to their energy/charge ($E/Q$) by a large field-of-view (≈$2\pi$) multi-slit collimator and deflection system with serrated plates and UV trap. The amount of deflection (and thus the incident angles) is measured precisely using the 2D position-sensitive start MCP assembly that detects secondary electrons (SEs) emitted from the carbon foil at the position of ion impact. The resolution achieved by this process is 5° in azimuth and elevation. The energy range is 0.05 – 20 keV/q with 5% resolution. The mass and mass-to-charge resolution are 10% and 8% respectively. The mass and power are 2 kg and 3.5 W.

The data products from the solar-wind instruments consist of distribution functions for solar-wind electrons and ions. Generally, these will be transmitted at degraded temporal resolutions, as moments in velocity space. However, high-resolution data are available for download per ground command. In addition, the SWI–SWE suite will provide rudimentary compositional information (He/H, Fe/O, He/O, $O^{7+}/O^{6+}$, $C^{5+}/C^{6+}$, $<Q_{Fe}>$). The highest time-resolution for a full spectrum is 5 seconds for electrons and ion measurement.

### 3.2.8 Energetic Particle Package

The instrument concepts described here provide measurements covering the energy range from ~20 keV/nucleon to ~100 MeV/nucleon for ions from H to Fe and covering ~0.02 to ~3 MeV for electrons. The EPP is based to a large extent on the SEP subsystem just launched on STEREO, also a 3-axis stabilized spacecraft (Figure 10).

The Low Energy Particle Telescope (LEPT) will measure the composition of energetic H to Ni ions from ≈3 to 100 MeV/nucleon, and electrons from ≈0.5 to 4 MeV, based on the LET sensors operating on STEREO (Mewaldt *et al.*, 2007). Each side of the double-ended LEPT has 5 silicon solid-state detectors (SSDs) arranged in 120° × 30° fan-shaped apertures. The 120° fans (2 of them) are in the ecliptic, one looking towards the Sun and one looking away. Centred between the two fans is a stack of rectangular SSDs that measure the position and residual energy of particles that stop in the stack or penetrate the entire stack. Pitch-angle distributions are measured over 240° of the ecliptic plane. The LEPT will measure energy spectra for species that include H, $^3$He,

$^4$He, C, N, O, Ne, Mg, Si, S, Ar, Ca Fe, and Ni. Overabundances of $^3$He, electrons, and heavy elements such as Fe are key signatures of particle acceleration in impulsive solar flares. Energy spectra can be measured on-board in ≈15 energy intervals per species. More than 1000 particles per second can be sorted into species *vs*. energy/nucleon matrices. LEPT will also identify ions with 30 ≤ Z ≤ 83 that are often enriched in impulsive SEP events. The geometry factor varies from 1.6 to 4 cm$^2$ sr depending on species.

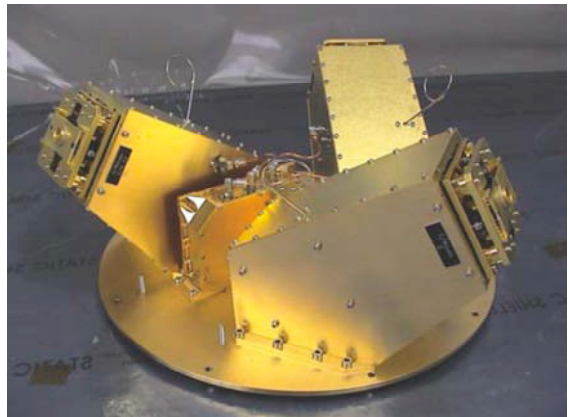

Figure 11: S/WAVES on STEREO from which the RPW could be derived

The Suprathermal Ion Telescope (SIT) telescope, based on an instrument that flies on STEREO (Mason *et al.*, 2007), uses the time-of-flight (TOF) *vs*. total energy method to identify ions from He to Fe with ≈20 keV/nucleon to ≈3 MeV/nucleon. Incident ions enter through a 1000 Å nickel foil, traverse a TOF region, and strike an SSD at the rear of the telescope, which measures the particle's kinetic energy. The combined TOF and energy signals can identify the mass and energy of abundant ions from He to Fe, including $^3$He. SIT is also sensitive to ions with 30 ≤ Z ≤ 83 and has a limited response to protons. The SIT looks 50° west of the Sun with a 20° × 20° field of view.

The Solar Electron Proton Telescope (SEPT) will measure the 3-D distribution of electrons and protons with good energy and time resolution. It is based on the SEPT sensors on STEREO (Muller-Mellin 2007), and includes two dual, double-ended magnet/foil telescopes that will identify electrons from 20 to 400 keV and protons from 20 keV to 7 MeV. The SEPT telescopes are constructed as open cylinders with two back-to-back solid-state detectors (SSDs), each 300 microns thick. A thin organic foil covers one aperture (electron side) while the other aperture is surrounded by a magnet (ion side). The thin foil leaves the electron spectrum unchanged, but stops low-energy protons. On the ion end a magnet sweeps away electrons, but lets ions pass and enter the SSD. Protons >0.4 MeV also enter the electron side. This contribution can be computed and subtracted from the electron spectrum using data from the ion aperture. Each dual SEPT sensor includes two parallel telescopes with the electron/ion sensors facing opposite directions to provide anisotropy information. The in-ecliptic sensor has a FOV that consists of four 52° cones, two looking toward the Sun and two looking away. The north-south sensor is mounted perpendicular to the ecliptic plane. The combined electron geometry factor is 0.5 cm$^2$sr while the ion geometry factor is 0.7 cm$^2$sr.

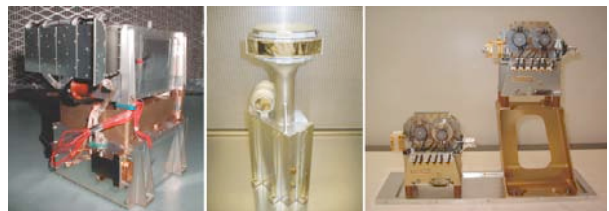

Figure 10: The SEP instrument suite of STEREO, from left to right, SIT, HET/LET, and SEP (Luhmann *et al.*, 2005)

The combined LEPT and SIT sensors will produce energy spectra for about a dozen species extending over ~20 energy intervals. Electron spectra from SEPT and LEPT will be accumulated in ~8 intervals. There will also be pitch-angle data for abundant species from LEPT for 16 directions.

### 3.2.9 Radio and Plasma Waves

The *Radio and Plasma Waves Analyser* (RPW) comprises two main sub-systems: a *Plasma Waves System* (PWS) covering *in-situ* measurements and the *Radio Astronomy Detector* (RASD) for remote sensing. The two sub-systems share some of the sensors and have common digital signal processing and interfaces with the spacecraft. The same receivers can be used to analyse the different types of waves detected by different sensors, for instance electric antennae and magnetic coils. All aspects of electromagnetic cleanliness are well understood and are no longer a problem for modern spacecraft designs (*e.g.* STEREO, see Figure 11).

The PWS will identify the *in-situ* plasma waves and kinetic modes comprising the electromagnetic part of the fluctuation and turbulence spectra. The PWS consists of two receivers: a *Thermal Noise Receiver* (TNR) and a *Low Frequency Receiver* (LFR) system. TNR is a very sensitive digital receiver designed specifically to measure the plasma thermal noise spectrum very accurately. Thermal noise spectroscopy gives very accurate measurements of electron density and temperature and is unperturbed by the large spacecraft floating potentials that afflict electron particle measurements. TNR will provide an accurate, and high time-resolution, reference for the solar wind experiments. The LFR is designed to study lower frequency, more intense, plasma waves associated with wind thermalization. The LFR will perform onboard processing of the electric and magnetic data and produce both spectra and waveforms to identify non-linear coherent structure. PWS can exchange burst triggers with other instruments in the *in-situ* payload to facilitate payload-wide burst modes. From *Helios* observations one can estimate that a sensitivity equal to $10^{-6}$ nT/Hz$^{1/2}$ will be required at 0.4 AU to identify unambiguously whether the observed waves are electromagnetic or electrostatic. The PWS will cover a broad band in frequencies, extending from about one Hz into the MHz range. Furthermore, multiple electric antennae allow for common-mode rejection of spacecraft generated noise and waves. The 3 components of the fluctuating magnetic field can be easily measured with a 3-axial search coil magnetometer arranged in a compact configuration and mounted on a short boom that should point in the anti-Sun direction. This boom (spacecraft item) is required for magnetic cleanliness reasons and could be shared with the MAG (see §3.2.6) as long as a minimum distance between the two sensors is respected.

The RASD will measure the solar and interplanetary radio waves in the frequency range from 100 kHz to 20 MHz, with a sweep period between 0.1 and 10 seconds and a high spectral resolution ($\Delta f/f \approx 0.07$). The RASD will observe plasma processes associated with energetic electrons from the corona up to about 0.5 AU. Since radio radiation is generally beamed (beam widths sometimes down to a few tens of degrees) more or less along a radial direction from the Sun, this technique is particularly relevant for different vantage points, for instance when POLARIS observes the far side of the Sun. The time history provided by the regular acquisition of the radio dynamic spectrum will help to synthesize the development of an active region. The time resolution required to detect the rapidly varying solar bursts varies with the radio frequency. This points to time resolution of the order of 0.1 seconds or better for the high frequencies and of 10 seconds for the low frequencies. The RASD can be of the classical super-heterodyne type. Frequency synthesisers will allow for a maximum flexibility in the choice of the observing frequencies (for instance to avoid "polluted" frequencies on the spacecraft). Each sub-receiver can consist of up to 256 selectable channels. Only a selection of these channels could be transmitted to the telemetry stream.

## 4 Mission and Spacecraft Design

The design of the POLARIS mission presented herein is based on the system baselines of the reference studies from Macdonald *et al.* (2006) [RD2] and from Liewer *et al* (2007) [RD1]. Similarly to the reference studies, the POLARIS mission employs a solar sail to arrive at the near-polar heliocentric science orbit. Both reference studies use the following methodology to transfer the science payload into a solar polar orbit

1. launch into an elliptical heliocentric orbit ($i = 0$) with positive $C_3$ (launch energy):
2. inward spiral phase to cranking orbit radius ($R_c$) (with residual non-zero inclination)
3. cranking phase from spiral phase end inclination to target science phase inclination:
4. possible third transfer phase during the science phase if orbital radius is greater than cranking phase

The baseline system for each document is in Table 3.

Table 3: Baseline Mission Parameters

|  | [RD1] | [RD2] |
|---|---|---|
| *Min Solar Approach Radius (AU) (= Cranking Radius)* | 0.48 | 0.48 |
| *Science phase inclination (degrees) w.r.t. Ecliptic* | 82.75 | 67.8 |
| *Launch injection mass (kg)* | 443 | 733 |
| *Sail mass (kg)* | 195 | 408 |
| *Payload mass (kg)* | 41 | 50 |
| *Platform mass (kg) (includes payload)* | 247 | 325 |
| *Sail side length (m)* | 153 | 178.9 |
| *Transfer duration (years)* | 5 | 6.7 |
| *Launch Vehicle* | Soyuz Fregat 2-1b | Delta IV |
| *Sail Areal Density (g/m$^2$)* | 9.298 | 14.16 |
| *Launch $C_3$ (km$^2$/s$^2$)* | 38 | 0.25 |

### *4.1 Mission Analysis*

This proposal presents a condensed set of results from mission analyses performed by Astrium UK for the POLARIS mission. The analysis draws from the two reference studies and incorporates a Venus Gravity Assist (VGA) to enhance the transfer performance and mass characteristics of the mission. It is believed that use of a VGA manoeuvre represents a more efficient and robust mission. The performance of a VGA mission is less sensitive to increases in spacecraft mass than the spiral-in equivalent. This is due to the fact that a very large mass (up to 1.4 tons with Soyuz) can be placed in an orbit with perihelion ≤ 0.48 AU and aphelion at ≈ 0.72 AU without requiring any thrust from the sail; by the time the sail is fully utilised (*i.e.* in the change of orbit inclination or *the cranking phase*), it is operating at a closer solar approach radius and is therefore more effective. This results in extra mass capability for relatively low performance deficit as presented here. There are however disadvantages to incorporating a VGA. The mission is limited to one of two sail deployment options:

- following Soyuz injection prior to VGA: perform VGA while fully deployed (potentially induce sail flexing)
- following VGA: sail must remain stowed for approximately 6 months (potential risk of cold-welding)

It is recommended that the sail be deployed following Soyuz injection to avoid cold welding. This will also require that the sail be used for the trajectory correction prior to the VGA, this will save hydrazine propellant but will require good sail control characteristics at an early stage in the mission. In the case that the spacecraft does not reach the desired VGA altitude, a time penalty may be added to the transfer duration to the science orbit.

A POLARIS Trajectory Simulator has been created to study the mission transfer from launch to science orbit. The VGA mission differs from the baseline mission with the following additional steps after Step 1 of the reference studies:

i. launch with a trajectory that intercepts Venus

ii. perform VGA and enter a post GA elliptical orbit with aphelion at Venus and a lower perihelion

iii. circularisation phase from post GA elliptical orbit to a circularised orbit with radius similar to perihelion of post GA elliptical orbit

The remaining Steps from 2 to 4 are identical to the reference studies. Figure 12 shows an example trajectory simulation output from the simulator (following VGA). The results are grouped in 3 scenarios as follows:

1. Scenario 1: implement VGA to baseline system of studies [RD1] and [RD2]: investigate usage of different launch vehicles
2. Scenario 2: optimize mass capabilities of launch vehicle with VGA
3. Scenario 3: investigate effects of heavier sail technology than that predicted in studies [RD1] and [RD2]

The sail design in the scenarios is based mainly on the system in RD2 as it represents a more pessimistic case with regard to sail technology standards incorporating a high sail areal density. The results of these scenarios are shown in Table 4 below; parameter values which differ from those of the corresponding studies are highlighted in red.

Table 4: Scenarios studied.

| Scenario | 1a | 1b | 1c | 2a | 2b | 2c | 3a | 3b |
|---|---|---|---|---|---|---|---|---|
| Reference System | RD1 | RD2 | RD2 | RD2 | RD2 | RD2 | RD2 | RD2 |
| Platform Mass (kg) | 443 | 325 | 325 | **601** | **656** | **707** | **539** | **467** |
| Sail side length (m) | 153 | 178.9 | **147** | **200** | 178.9 | 178.9 | 178.9 | 178.9 |
| Cranking Radius (AU) | 0.48 | 0.48 | 0.48 | 0.48 | **0.4** | 0.48 | 0.48 | **0.4** |
| Trip Time (years) | **5.5** | **5.51** | **6.74** | 6.7 | 6.7 | **8.23** | **8.23** | **6.39** |
| Sail Areal Density (g/m$^2$) | 9.298 | 14.16 | 14.16 | 14.16 | 14.16 | 14.16 | **20** | **20** |
| Launcher | **Dnepr** | **Soyuz Fregat** | **Dnepr** | **Soyuz Fregat** | **Soyuz Fregat** | **Soyuz Fregat** | **Soyuz Fregat** | **Soyuz Fregat** |

The POLARIS baseline is Scenario 2c (solar sail system of [RD2] and Soyuz Fregat launcher).

In Scenario 1, it was shown that smaller, lower-cost launchers could be used whilst still giving comparable performance to the baseline studies. The most cost-effective case for [RD2] is scenario 1c where the Dnepr is used: this is accompanied by a decrease in sail size and a corresponding increase in trip time. This scenario represents a low-cost solution however, it allows very little room for mass evolution, *i.e.* the platform mass is restricted to the 325 kg of the [RD2] baseline. Furthermore, the smaller launcher fairing of the Dnepr may lead to kick-stage/sail canister accommodation issues (although this may be somewhat alleviated by the decreased sail size).

Scenario 2 looks at utilising the full launch mass capabilities of the VGA mission by allowing greater mass to be inserted into the science orbit, this reduces the risk of such a mass-sensitive mission and gives the option to accommodate a heavier science payload. Two approaches were used to retain the trip time performance with the increased mass:

- scaling the sail size
- scaling the cranking radius.

With the Soyuz mass fully utilised, a sail size of 203 × 203 m$^2$ (Scenario 2a) will allow for a platform mass of 587 kg. Alternatively, reducing the cranking radius to 0.4 AU (Scenario 2b) will allow for a platform mass of 641 kg. Both cases represent a large increase over the [RD2] baseline platform mass of 325 kg. Scenarios 2a and 2b represent a higher-cost mission than scenario1c due to the need for a Soyuz launcher; however the extra mass margin available for the platform/payload/sail represents a lower-risk mission regarding technology standards. If the sail design and cranking radius are to be kept identical to the baseline [RD2] system (scenario 2c), the platform mass increases to 707 kg with a trip time penalty of 1.74 years. This solution may be desirable if the reduction of mission duration is seen to be of low priority. Scenario 2c (highlighted in dark grey) is chosen as the POLARIS baseline mission design.

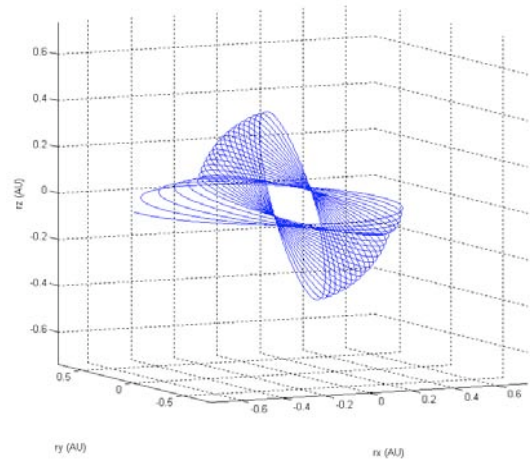

Figure 12:. An example of a post VGA transfer with a cranking radius of 0.4AU (cranking to 75° w.r.t. the solar equator)

Scenario 3 quantifies the effect of heavier sail technology. This shows that at the predicted worst-case Sail Areal Density of 20 $g/m^2$, the VGA mission will allow a platform mass of 539 kg with the reduced performance of just over 8 years trip time. This scenario gives confidence in the robustness of the mission to cope with adverse technology issues.

The science orbit has been chosen at 75° inclination to the solar equator for the reason stated in [RD2] that ≈ 30 % of the orbit will be spent above a heliographic latitude of 60°. As the Sun is inclined at 7.2° with respect to the ecliptic, the science orbit must be oriented such that the Right Ascension of the Ascending Node (RAAN) is aligned with the point at which the solar equator passes through the ecliptic plane. This alignment may be achieved through careful selection of the launch epoch and tuning of the steering law used in the circularization phase. Note that the RAAN of the POLARIS orbit precesses by approximately 40° during the circularization phase; the precession rate can be controlled through modification of the steering law.

*4.2 Platform Configuration*

The POLARIS platform can be constructed through large-scale re-use of the SolO platform. The SolO environment is more severe than that of POLARIS; therefore, the SolO-derived platform represents a worst-case POLARIS platform design. For means of technology re-use, the SolO platform structure is directly implemented into the POLARIS platform. The SolO structure is designed to accommodate a similar set of science payload instruments including a Visible Imager and Magnetograph, EUV Spectrometer, EUV Imager, Visible Coronagraph and an Energetic Particle Detector. The SolO Payload element dimensions are comparable to those of the POLARIS payload allowing easy integration. However, the SolO hydrazine tanks are much larger than those required for POLARIS; therefore the SolO derived structure represents the worst-case. Potential mass savings may be made through reduction of the panel dimensions.

Due to entirely different trajectory characteristics and the implementation of a solar sail, the POLARIS propulsion system will use a much lower proportion of hydrazine compared to that of SolO. The SolO hydrazine propellant for the 2015 launch baseline is greater than 150 kg and employs 12 × 10 N thrusters. The POLARIS propulsion system also utilises 12 × 10 N thrusters; however, the expected hydrazine mass is estimated at worst-case of 30 kg; this is based on scaling the [RD1] propellant mass by the ratio of flight system masses with an additional 10 kg for any potential VGA trajectory corrections (non-sail) required. It is assumed that the sail will mainly be used to correct the spacecraft trajectory prior to the VGA. The POLARIS baseline propulsion system is therefore based on the SolO propulsion system with appropriately scaled hydrazine tanks.

*4.5 Telemetry, Tracking and Command*

The POLARIS mission requires a larger science data downlink than the SolO mission and therefore the SolO communications system will not be re-used. Instead, the higher performance communication system of the *Mercury Polar Orbiter* (MPO) element of *Bepi-Colombo* can be employed; as the SolO communications system has direct heritage with that of the MPO, integration into the POLARIS platform is expected to be straightforward. The POLARIS communications system comprises of Low Gain, Medium Gain and High Gain antennae (LGA, MGA, HGA), and their corresponding support mechanisms.

Uplink/downlink throughout the cruise phase can be performed using an optimal combination of the HGA, the MGA (350bps at 1.7AU) and the LGAs (~8bps at 1.56AU), all providing hemispherical coverage (HGA and MGA are 2-axis pointed).

The MPO communications system of *Bepi-Colombo* has a larger data downlink rate than SolO communications system. The total data downlink rate is 260 kbps at 1 AU and 200 kbps at 1 AU, for *Bepi-Colombo* and SolO, respectively. The Science Payload aggregate data rate is approximately 86.6 kbps including 20% overhead instrument housekeeping and lossy compression of a factor of 4 (173.2 kbps for lossless compression). In order to downlink this data during the science phase of the mission a single downlink of 8 hours must be performed every day. Science data downlink is performed in both X-band and Ka-band. The dataflow analysis assumes the following.

- Downlink via HGA is always possible throughout science mission
- No restriction on antenna pointing angle
- No space-segment communications outages
- Downlink data rate simply scaled to inverse-square law with solar distance (calibrated at 1AU)

Simultaneous X-band/Ka-band downlink operations can be performed in conjunction with the Cebreros ESA ground station: this will only be possible if the SolO mission does not coincide with the *Bepi-Colombo* Science Phase.

Potential sources of communications outage for example, interference of the physical sail on the HGA signal or limitations of the HGA pointing range have been identified. Potential resolutions of this problem may be to upgrade the communications system to increase telemetry rate or to augment the downlink schedule with extra downlinks at an alternative station such as New Norcia

The MPO communications system gives comparable performance to the systems of [RD1] and [RD2].

### 4.6 Data Handling

It is expected that the on-board processing capabilities required for the POLARIS Science Phase will be less stringent that that of SolO as no reduction of data volume is performed on the space segment other than compression by a ratio of 4. The POLARIS Payload records data at a constant rate of 86.6 kbps (effective rate, after lossy compression) during the Science Phase, with the communications subsystem described in Section 2.2, this requires an onboard data storage capacity of no greater than 300 Gbits which is within the reused SolO Data Handling System End-Of-Life capacity of 360 Gbits.

### 4.7 Attitude and Orbit Control System

The 3-axis stabilized SolO AOCS subsystem can be completely re-used for the POLARIS Platform. The SolO Relative Pointing Accuracy of 1 arcsec over 10 seconds is slightly tighter than that of [RD2] (0.44 arcsec over 3 seconds). Similarly the SolO Absolute Pointing Accuracy of 10 arcsec matches the equivalent of [RD2]. The POLARIS Platform AOCS system consists of Sun acquisition sensors, star trackers, Astrix 200 fibre-optic gyros and reaction wheels. Attitude estimation is based on hybridisation of an autonomous star tracker and the gyros. The Sun sensors may be used for attitude acquisition and safe modes. The reaction wheels are used in conjunction with the 12 N thrusters for actuation. With these performances only the DSI and EUVI instruments will require internal active pointing.

### 4.8 Thermal Control

The POLARIS Platform thermal control subsystem consists of a modified version of the SolO thermal control subsystem. SolO is designed to reach a minimum solar approach radius (MSAR) less than 0.3 AU and therefore requires a heat shield as well as radiators. The POLARIS mission presented in this study is designed to reach an MSAR of no less than 0.4 AU, at this MSAR, the heat shield can be replaced with a lighter thermal blanket.

This analysis assumes that the thermal blanket covers the same area as the original SolO heat shield of $\approx 5.1$ m$^2$; this ensures a comparable thermal environment for other SolO inherited subsystems.

20-Layer Aluminium and Silver coated FEP at 2 kg/m$^2$ may be employed which have good thermal properties with a large margin between the steady-state temperature and the maximum operating temperature. Either solution can be employed at both 0.48 and 0.4 AU and both have good electrical conductivity properties. The heavier Aluminium/Silver FEP has been selected as a baseline.

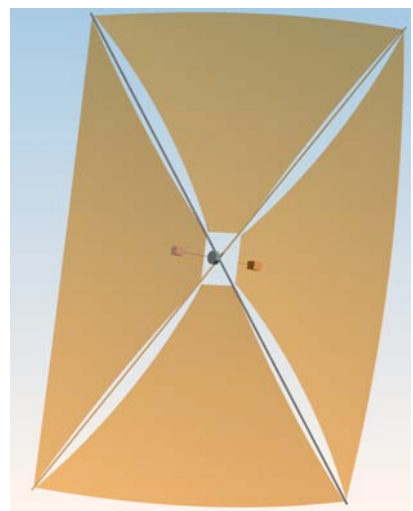

Figure 13: A view of the spacecraft with the sail deployed on its mast. The sail size is a square of 180 m while the spacecraft fits into a 2-m on a side cube (View courtesy of J.-C.Leclec'h).

### 4.9 Power

For the POLARIS baseline, the SolO solar panel area of 5.6 m$^2$ can be scaled down to the 1.36 m$^2$ solar panel area of [RD2] to reduce the mass of the power subsystem; this power system satisfies the power requirements at MSAR of both 0.4 AU and 0.48 AU (based on payload 1.5 watts/kg). The remainder of the SolO power subsystem including batteries and support mechanisms is left unaltered in order to promote technology re-use.

### 4.10 Mass and Power Budgets

The total POLARIS dry mass (worst case) is estimated to be 583 kg. Adding a hydrazine propellant mass of 30 kg gives a platform wet mass of 613 kg, which is compatible with the 707 kg platform wet mass limit of Scenario 2c leaving a comfortable margin of ≈90 kg. The total launch mass including system margins of 20% is 1224 kg (to be compared with the Soyuz capability of 1350 kg for an MSAR of 0.48 AU). The worst-case power budget for the POLARIS mission is 924 Watts (including system margin of 20%), which is expected to be that of the Science phase.

### 4.11 Cleanliness program

The Electric and Magnetic Cleanliness (EMC) requirements are driven by the RPW and MAG instruments, that carry an antenna and a magnetometer, respectively. As for magnetic issues, the use of permanent magnets and soft magnetic materials, and careful design of power distribution systems to minimise current loops is quite straightforward. As for electrical issues, several measures can be taken to ensure that the POLARIS spacecraft is clean from the point of view of both conducted and radiated electromagnetic interference.

Aboard POLARIS, there are several UV and EUV instruments requiring specific hydrocarbon cleanliness measures, and optical instruments requiring particulate cleanliness measures similar to those taken for the SOHO and STEREO mission. A similar cleanliness program will be put in place for the SolO mission.

### 4.12 Sail Design and Considerations

The POLARIS baseline draws on the Bi-State Gimbal baseline sail design of RD2 with a CP-1 sail and an areal density of 14.16 g/m$^2$. In this design, the control of the sail orientation and thrust vector is achieved by offsetting the centre of solar radiation pressure from the centre of mass of the system using a bi-state (locked or unlocked) gimballed to connect the sail system to the spacecraft (Figure 13). Unlocking the gimbal allows the position of the platform relative to the sail to be adjusted. The offset produces a net torque on the sail that can be adjusted by changing the relative position of the Platform. The sail truss design is based on the ATK-Able Coilable® Masts scheduled to be flown on the New Millenium Program's Space Technology 8 in 2008 (McEachen et al, 2005). Using this sail design for POLARIS gives a realistic areal density and provides a good level of technology heritage.

The sail temperature at 0.4 AU (Scenario 2b, worst-case) reaches 200°C [RD2], which is comfortably within the 260°C continuous operating temperature limit of the CP-1 sail material.

Several critical issues must be investigated in future work when applying the [RD2] sail system to the POLARIS baseline platform. A platform mass of > 600 kg will transfer larger forces through the control mast of the sail system; structural analysis must be performed.

The development of the sail could be done according to the following model philosophy:

- Engineering Model: boom and sail development equipment
- Qualification Model: boom, sail and all deployment mechanisms)
- Flight models: 4 booms, 4 sails (sail made of 4 elements)
- Flight spare: one boom, one sail

This development approach has been used for costing the sail system.

### 4.13 Mission operations

The management of the operations will be similar to that of the *Ulysses* mission. The spacecraft will be operated at the NASA-funded Mission Operations Centre (MOC). The scheduling of operations, spacecraft monitoring and navigation will be done by JPL, while the solar sail control software will be provided by ESA through ESOC. During the cruise phase, the *in-situ* instruments will be operated provided that they do not interfere with the solar sail operation. The commissioning of the remote sensing instruments will be done at the beginning of the

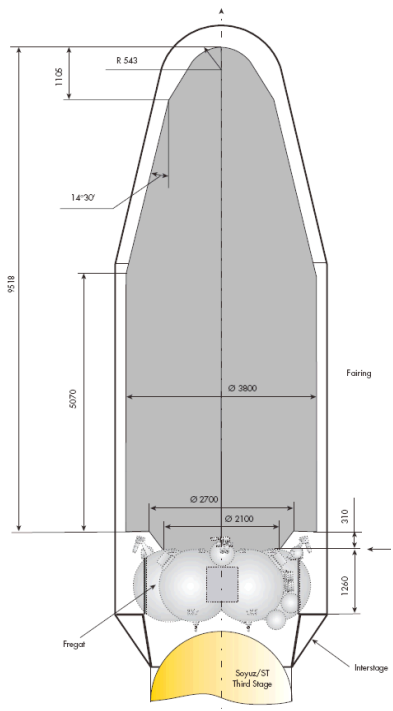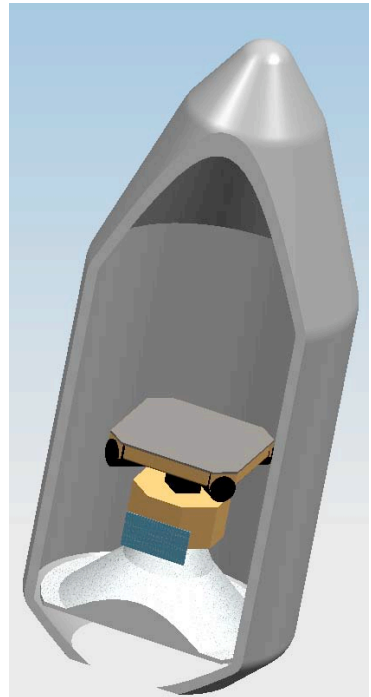

Figure 14: View of the spacecraft with the solar sail in launch configuration in the 4-m diameter Soyuz fairing. Spacecraft dimensions are derived from [RD2] (View courtesy of J.-C.Leclec'h).

Science orbit.

The 35-m Cebreros antenna will be used for communications with the spacecraft. Thanks to its dual frequency capabilities, it offers at the time of writing than the DSN antennae of similar size.

### *4.14 Launcher requirement*

One of the main requirements on the launcher is imposed on the fairing, which should be at least 3 m diameter. Figure 14 shows the spacecraft in the new Soyuz fairing of 4-m diameter.

## 5 Science Operations and Archiving

Science planning and operations are relatively simple for POLARIS since the spacecraft always remains Sun-pointed once the science orbit is reached. Only data rate allocations will vary on a week-by-week schedule, depending on the science goals for the particular period. This allows for optimization of the science return. Schedules would be prepared weekly or less frequently as needed and delivered to the NASA MOC to be turned into uplink command sequences. The planning cycle would be similar to that used for the STEREO mission.

The MOC will be responsible for providing the Level 0 data streams to each instrument team for processing. The PI shall be responsible for providing the fully calibrated science data (Level 2 products) derived from the engineering data (Level 1 products), themselves made from the raw data (Level 0). The anticipated volume of data for the 2-year mission will be 25 Tbytes for all levels of data, stored in the long-term archive. There will be two Science Data Centres (SDCs), one in Europe (at the ESAC of Villafranca) and one in the US (at JPL). The processed science data will be collected by the POLARIS science data centres for distribution and archiving.

We envision that the Level 2 data products will be freely available to the science community as soon as made available by the teams. Such a data policy is now common amongst solar missions such as STEREO, SDO, and *Hinode*.

## 6 Key technology areas

### *6.1 Payload TRL and technology development*

For most of the instruments, there is a concept that has already been flown in a space environment. The Technology Readiness Level (TRL) for these conceptual instruments is 9. Several other concepts are either working in the laboratory, or are in preparation for balloon-borne mission or a sounding rocket. The TRL for these latter concepts is 6. In a latter case, technology development used by *Bepi-Colombo* and SolO are expected for low power consumption detectors such as APS, liquid crystal variable retarders (LCVR), and a solid etalon. It is anticipated that most of the technological challenge shall not be on the payload but on the spacecraft item – the solar sail.

## 6.2 Mission and spacecraft technology challenges

The use of a solar sail has been thought of as early as 1924 by Tsander. Since then there have been various attempts at flying solar sail demonstrators. It is clearly identified that the use of solar sail is the major challenge of the POLARIS mission. Developments are required in the following areas:

- Solar Sail Material
- Solar Sail Deployment (booms)
- Solar Sail Attitude and Orbit Control system
- Solar Sail Jettison Mechanism
- Communications and AIV

The Technology Readiness Level is 3 to 4 for spacecraft items. All of these items are being studied in the *GeoSail* Technology Reference Study. *Geosail* is a small 250-kg spacecraft using a 50 × 50 m$^2$ sail in an elliptical orbit around the Earth.

## 6.3 Mission flexibility

The POLARIS mission has been designed to cope with unforeseen deviations from the predicted sail technology standards. The current baseline incorporates a simple platform constructed through extensive re-use of the SolO and *Bepi-Colombo* platform elements; this platform design currently incorporates only the most fundamental mass-saving design changes and as a result embodies a relatively low cost and low risk solution. Although the baseline platform is of relatively high-mass, the mission analysis presented herein shows that a feasible mission can still be accomplished with a trip time to science orbit of less than 8.5 years. This reasonable trip time along with a large spacecraft wet mass and low-cost Soyuz launch is made possible through the use of the VGA.

Were the sail technology to become heavier than expected for a given sail size, design refinements to the baseline platform could then be made accordingly; thereby, allowing the simple transfer of the allocation of mass budget from the platform directly to the sail subsystem. This process would retain the same overall mass and characteristic acceleration and hence trip time performance. The above approach of mass transfer through platform design refinement can be thought of as a safety net for the case of adverse sail technology standards. The main candidate platform elements for such refinement are as follows:

- *Structure:* panels may be reduced in size due to smaller hydrazine tanks; potential reconfiguration of payload and subsystem units required
- *Communications system*: the re-used *Bepi-Colombo* antennae are designed to operate at <0.3AU; the 0.48 AU POLARIS baseline mission may allow for lighter antennae designs which operate at lower temperature
- *Solar Panels:* The SolO solar arrays are designed to operate at < 0.3 AU; the 0.48 AU POLARIS baseline mission may allow for lighter array designs which operate at lower temperature
- *Propellant:* further mission analysis may be done to reduce the baseline 30 kg of propellant

Platform refinement may also be undertaken in order to reduce the trip time, hence saving on operating costs.

## 6.4 Non-phased Orbits and Varying Solar Approach Radii

Missions scenarios presented herein look at two different minimum solar approach radii 0.48 AU and 0.4 AU corresponding to the resonant science phase orbits N = 3 and N = 4 with the Earth orbital period (derived from N$^{-2/3}$AU, see RD1). It has been identified that the solar polar specific science objectives do not specifically require a resonant orbit. In fact, the only driver on orbit resonance is communications operations during the science phase. A non-phased orbit will generally incorporate some amount of communications outages throughout the science phase due to solar radio interference during sail/Earth/Sun conjunctions; however, such outages are have been shown to occur for less than 3% of the science phase for the range of solar distances expected for POLARIS for a very pessimistic solar interference half-cone of 5$^o$ (from Earth's perspective). Such low outage statistics suggests that it is not necessary to restrict the POLARIS science orbit to a phased solution unless the communications system is operating to the very limit of its capabilities.

With the requirement of a phased orbit removed, the minimum solar approach radius and therefore the cranking orbit radius may be chosen to give optimal performance. Generally, a lower cranking radius will give more favourable injection mass/trip time performance characteristics with the optimal value at approximately 0.35 AU; beyond this point, the injection mass is prohibitively restricted due to excessive $C_3$ required for the VGA manoeuvre. From this we may consider that the POLARIS system may be optimized through reduction of the cranking radius.

The main driver of the minimum cranking radius is the thermal effect on the platform, and sail. Due to the near-solar nature of the inherited technology, very little changes need be made to the baseline POLARIS platform in order to operate at solar distances of less than 0.48 AU. It has been shown in Section 4.8 that using a thermal blanket, the platform can comfortably withstand the thermal environment at 0.4 AU. It has been shown in RD5 that the baseline sail system material can also operate at 0.4 AU at a temperature of 200ºC.

# 7 Preliminary programmatics

## 7.1 Mission management structure

ESA will be responsible for procuring the spacecraft, the solar sail system, the launch, the ground stations, and the European Science Data Centre, located at ESAC. NASA will be responsible for procuring and operating the MOC, navigation and the US Science Data Centre. All payload instruments and data processing will be provided by National Agencies of ESA Member states or funded by NASA.

## 7.2 Mission Schedule Drivers

The development of the solar sail technologies is the obvious main driver for the starting of the mission implementation. It is foreseen that necessary in-flight demonstration of these technologies will be required before proceeding to the implementation phase. A mission such as *Geosail* could serve as a technology demonstration mission.

### Reference documents

[RD1] Liewer, P.C. *et al.*, 2007, *Solar Polar Imager: Observing Solar Activity from a New Perspective, NASA Vision Mission Study Final Report,* JPL D-33704 Rev. A.

[RD2] Macdonald, M. *et al.*, 2006, *Journal of Spacecraft and Rockets*, **43(5)**, 960

[RD3] SolO Heat Shield/System Technology Study, SOL-T-ASTR-PRS-29

[RD4] *Bepi-Colombo* System Budgets Report, BC-ASD-TN-00016

[RD5] Scalable Solar Sail Subsystem Design Considerations AIAA 2002-1703